\documentclass[preprint,aps,12pt,showpacs,nofootinbib,tightenlines]{revtex4}
\usepackage{amsmath}
\usepackage{amssymb}
\usepackage{slashed}
\usepackage{epsfig}
\usepackage{graphicx}
\usepackage{enumerate}
\usepackage{multirow}
\usepackage[dvips,usenames]{color}
\begin{document}

\newcommand{\beq}{\begin{eqnarray}}
\newcommand{\eeq}{\end{eqnarray}}
\newcommand{\non}{\nonumber\\ }

\newcommand{\mb}{m_B }
\newcommand{\mtp}{m_{t'}}
\newcommand{\mw}{m_W }
\newcommand{\im}{{\rm Im} }


\def \cpc{ {\bf Chin. Phys. C} }
\def \csb{ {\bf Chin. Sci. Bull.} }
\def \ctp{ {\bf Commun. Theor. Phys. } }
\def \ijmpa{ {\bf Int. J. Mod. Phys. A}  }
\def \copc{ {\bf Comput. Phys. Commum. } }
\def \epjc{{\bf Eur. Phys. J. C} }
\def \jpg{ {\bf J. Phys. G} }
\def \npb{ {\bf Nucl. Phys. B} }
\def \plb{ {\bf Phys. Lett. B} }
\def \pr{  {\bf Phys. Rep.} }
\def \prd{ {\bf Phys. Rev. D} }
\def \prl{ {\bf Phys. Rev. Lett.}  }
\def \ptp{ {\bf Prog. Theor. Phys. }  }
\def \rmp{ {\bf Rev. Mod. Phys. }  }
\def \zpc{ {\bf Z. Phys. C}  }
\def \jhep{ {\bf JHEP}  }
\def \epl{ {\bf EPL} }
\def \appb{ {\bf Acta Phys. Polon. B} }
\title{Searches for the FCNC couplings from top-Higgs associated production signal with $h\to \gamma\gamma$ at the LHC}
\author{Yao-Bei Liu$^{1}$\footnote{E-mail: liuyaobei@sina.com}, Zhen-Jun Xiao$^{2}$\footnote{E-mail: xiaozhenjun@njnu.edu.cn}}
\affiliation{1. Henan Institute of Science and Technology, Xinxiang 453003, P.R.China \\
2. Department of Physics and Institute of Theoretical
Physics, Nanjing Normal University, Nanjing 210023, P.R.China }

\begin{abstract}
In this paper, we study the observability of the top-Higgs flavor changing neutral current (FCNC) $tqh$ coupling through the process $pp\to qg\to t(\to \ell^{+}b\nu)h(\to \gamma\gamma)$ at the Large Hadron Collider~(LHC), where $\ell=e, \mu$. Our numerical results show that, in some parameter regions, the LHC may observe the above signals at the $5\sigma$ level. Otherwise, the branching ratios $Br(t\to uh)$ and $Br(t\to ch)$ can be respectively probed to $0.036\%$ and $0.13\%$ at $3\sigma$ level at 14 TeV LHC with the high integrated luminosity of 3000 fb$^{-1}$. On the other hand, studying the charge ratio for the lepton in top quark decay can be not only used to
discriminate between signal and backgrounds, but also used to discriminate between $tuh$ and $tch$ couplings, for which anomalous single top production comes from the up initiated channel and charm initiated channel.
\end{abstract}

\maketitle

\newpage
\section{Introduction}
The discovery of a Higgs boson with a mass about 125 GeV is the undisputed highlight of Run-I of the CERN's Large Hadron Collider (LHC)~\cite{atlas,cms}. So far the measured couplings of the Higgs boson with fermions and gauge bosons are found to be in agreement with the predictions of the Standard Model (SM)~\cite{p1,p2}. A major target of the future LHC programme is to study the intrinsic properties of the discovered Higgs boson. In view of the large top quark mass and the large number of top quarks produced at the LHC, it is attractive to investigate the anomalous flavor-changing neutral current (FCNC) top-Higgs couplings~\cite{t1,t2,t3,t4,t5,t6,t7,t8,t9}.

In the SM, the FCNC couplings in the top sector are strongly suppressed due to
the Glashow-Iliopoulos-Maiani (GIM) mechanism~\cite{gim}, which can only occur at loop-level with the expected branching ratios of order about $10^{-15}-10^{-12}$~\cite{gim1,gim2}. However, in many new physics (NP) models beyond the SM, such as the Minimal Supersymmetric Standard Model~(MSSM)~\cite{susy1,susy2,susy3,susy4,Balazs:1998sb}, Two-Higgs-Doublet Models~(THDM)~\cite{2hdm-1,2hdm-2,2hdm-3,2hdm-4,He:2002fd}, Extra Dimensions~(ED) \cite{ED}, Little Higgs Models~(LHT)~\cite{lht}, and the other miscellaneous models~\cite{other1,other2,other3,other4,He:1999vp}, some FCNC processes involving the top quark can be greatly enhanced by extending the flavor structures, which makes them potentially accessible at current and future high-energy colliders.
Thus, any signal for top quark FCNC process at a measurable rate would be a robust evidence for NP. Since we do not know which type of NP models will be responsible for the possible deviation, it is better to study these processes with a model-independent method. So far, there are already many studies on the probe of the anomalous FCNC couplings in the top quark sector within model-independent method~\cite{tfcnc-th,multilepton,tfcnc-exp,f1,f2,f3,f4}.

Recently, the ATLAS and CMS collaborations have set the upper limits of $Br(t\to qH)<0.79\%$~ \cite{atlas-fcnc} and $Br(t\to cH) < 0.56\%$ at $95\%$ confidence level (C.L.)~\cite{cms-fcnc}.
The production of the top pair ($t\bar{t}$) and associated top-Higgs ($th$) via FCNC couplings has been emphasized in the recent studies~\cite{prd86-094014,jhep-1407-046,prd89-054011,13112028,wlei-jhep,prd92-074015,160204670,prd92-113012,th-3,th-4,th-5,th-6}.
Especially, the author of~\cite{prd86-094014} studied the anomalous production of $th$ via the FCNC interaction of $tqh$ at the LHC through the $h\to b\bar{b}$ channel including complete QCD next-to-leading order (NLO) corrections. The anomalous production of $th$ at the LHC
originating from FCNC interactions in $tqg$ and $tqh$ vertices has also been studied via the $h\to b\bar{b}$ decay channel~\cite{prd89-054011}. Although the branching ratio of Higgs diphoton decay channel is small, it has the advantages of good resolution on the Higgs mass and
small QCD backgrounds. Thus in this paper, we mainly investigate the top-Higgs FCNC interactions through $pp \to th$ with the sequent decays $t\to W^{+}b \to \ell^+ \nu b$ and $h \to \gamma\gamma$ at the LHC.

This paper is organized as follows. In Sec.~II, we give a brief introduction to the anomalous FCNC $tqh$ couplings and our selected production channel. In Sec.~III, we discuss the observability of the top-Higgs FCNC couplings through the process $pp \to t(\to W^{+}b\to \ell^{+}\nu b)h(\to \gamma\gamma)$ at 14 TeV LHC. In Sec.~IV, we discuss the leptonic charge ratio of the signal and backgrounds. Finally, we summarize our conclusions in Section.~V.

\section{Calculation framework}
\subsection{Top-Higgs FCNC couplings}
In general, the effective Lagrangian describing the FCNC Yukawa interactions of a light up-type quark with the top quark and a Higgs boson can be written as
~\cite{plb703-306}
\begin{equation}
{\cal L}= \lambda_{tuh}\bar{t}Hu+\lambda_{tch}\bar{t}Hc+h.c.,
\label{tqh}
\end{equation}
where the real coefficient $\lambda_{tqh}$ $(q=u,c)$ denotes the strength of the top-Higgs FCNC coupling. At the leading order (LO) and the NLO, the decay widths of the dominant top quark decay mode $t\rightarrow Wb$ could be found in Ref.~\cite{twb}. After neglecting all the light quark masses and assuming the dominant top decay width $t \to bW$, the branching ratio of $t \to qh$ can be approximately given by \cite{jhep-1407-046}:
\begin{equation}
Br(t \to qh) = \frac{\lambda^{2}_{tqH}}{\sqrt{2} m^2_t G_F}\frac{(1-x^2_h)^2}{(1-x^2_W)^2 (1+2x^2_W)}\kappa_{QCD} \simeq 0.58\lambda_{tqh}^{2},
\end{equation}
with the Fermi constant $G_F$, the top quark mass $m_{t}$, the $W$ boson mass $m_W$, the Higgs mass $m_h$ and $x_i=m_i/m_t~(i=W,\ h)$. Here the factor $\kappa_{QCD}$ is the NLO QCD correction to $Br(t \to qh)$ and equals about 1.1~\cite{nlo}.

Currently, the stringent constraints on the anomalous FCNC couplings are set exploiting the experimental data of the ATLAS and CMS Collaborations \cite{atlas-fcnc,cms-fcnc}.
On the other hand, the low energy observables, such as $D^{0}-\bar{D^{0}}$ mixing~\cite{prd81-077701} and $Z\to c\bar{c}$~\cite{prd72-057504} can also be used to constrain
the top quark flavor violation in the $tqH$ vertex. With the 125 GeV Higgs boson mass, upper
limits of $Br(t\to cH) < 2.1 \times 10^{-3}$ have been obtained from the $Z\to c\bar{c}$ decay~\cite{prd92-113012}. The author of~\cite{jhep06-033} also derived model-independent constraints on the $tcH$ and $tuH$ couplings that arised from the bounds on hadronic electric dipole moments.

\subsection{The production processes}
\begin{figure}[htb]\vspace{0.5cm}
\begin{center}
\centerline{\epsfxsize=12cm \epsffile{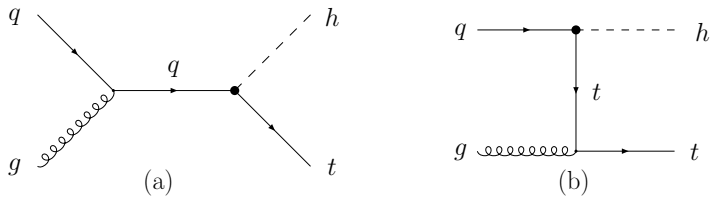}}
\vspace{-12cm}
\caption{The LO Feynman diagrams for $th$ associated production at the LHC through FCNC top-Higgs interactions. Here $q=u,c$.}
\label{qg}
\end{center}
\end{figure}

At the LHC, the parton level signal process at the tree-level via the FCNC $htq$ couplings can be expressed as
\beq
qg\to tH,
\eeq
where $q$ is $u$ or $c$ quark. The Feynman diagrams are shown in Fig.~\ref{qg}. Obviously, the conjugate process $\bar{t}+h$ production can also occur at the tree level.

We first implement the $tqH$ FCNC interactions by using the \texttt{FeynRules} package~\cite{feynrules}. The LO cross section are computed using \texttt{MadGraph5-aMC$@$NLO}~\cite{mg5} with CTEQ6L parton distribution function (PDF)~\cite{cteq}, setting the renormalization and factorization scales to be $\mu_R=\mu_F=\mu_0/2=(m_t + m_h)/2$. In this work, we assume $\lambda_{tqh}\leq0.1$ to satisfy the direct constraints from the ATLAS and CMS results~\cite{atlas-fcnc,cms-fcnc}. The SM input parameters are taken as follows~\cite{pdg}:
\begin{align}
m_H&=125{\rm ~GeV}, \quad m_t=173.21{\rm ~GeV}, \quad m_W=80.385{\rm ~GeV},\\ \nonumber
\alpha(m_Z)&=1/127.9, \quad \alpha_{s}(m_Z)=0.1185, \quad G_F=1.166370\times 10^{-5}\ {\rm GeV^{-2}}.
\end{align}

\begin{figure}[b]
\begin{center}
\vspace{-0.5cm}
\centerline{\epsfxsize=13cm \epsffile{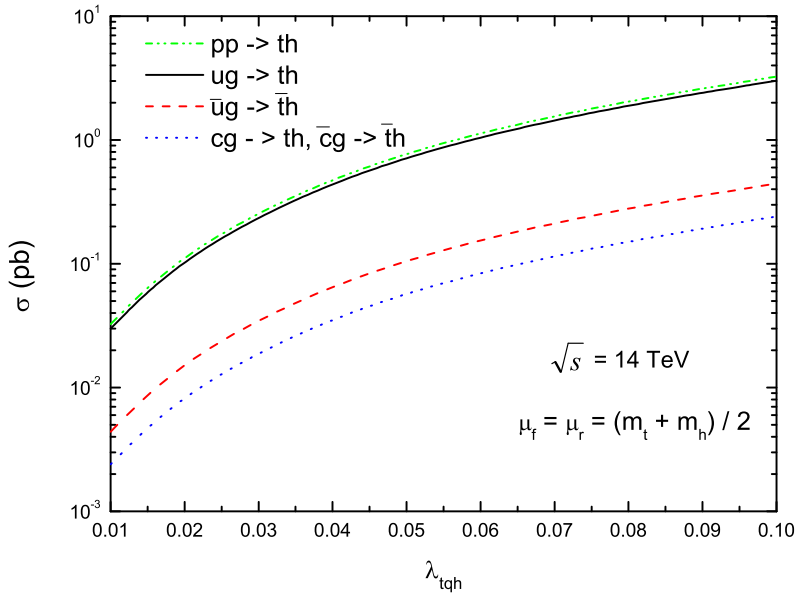}}
\caption{The dependence of the cross sections $\sigma$ at 14 TeV LHC on the top-Higgs FCNC couplings $\lambda_{tqh}$.}
\label{cross}
\end{center}
\end{figure}
In Fig.~\ref{cross}, we show the dependence of the cross sections $\sigma_{th}$ and $\sigma_{\bar{t}h}$ on the top-Higgs FCNC couplings $\lambda_{tqh}$ at 14 TeV LHC for different processes. From Fig.~\ref{cross}, we can see that the cross section of $ug \to th$ is larger than that for other processes. To be specific, when $\sqrt{s}= 14$ TeV and $\lambda_{tqh}=0.1$, the production cross section $\sigma$ is about 20 pb, which is roughly an order of magnitude larger than that for the conjugate process $\bar{u}g\to \bar{t}h$ due to the difference between the $u$-quark and $\bar{u}$-quark PDF of the proton.
Thus, for a given center-of-mass (c.m.) energy and luminosity, more leptons will be observed than anti-leptons considering the leptonic top decays $t\to W^{+}(\to \ell^{+}\nu_{\ell})b$ and $\bar{t}\to W^{-}(\to \ell^{-}\bar{\nu}_{\ell})\bar{b}$.
On the other hand, since the $c$-quark and $\bar{c}$-quark have the similar small PDF,
 the production rates of top and anti-top quarks
from the processes of $gc(\bar{c})\to ht(\bar{t})$ are almost the same and smaller than that for the process $ug \to th$.  This implies that the sensitivity to the coupling $\lambda_{tuh}$ will be better than $\lambda_{tch}$.

\section{Signal and discovery potentiality}
\begin{figure}[htb]\vspace{0.5cm}
\begin{center}
\centerline{\epsfxsize=16cm \epsffile{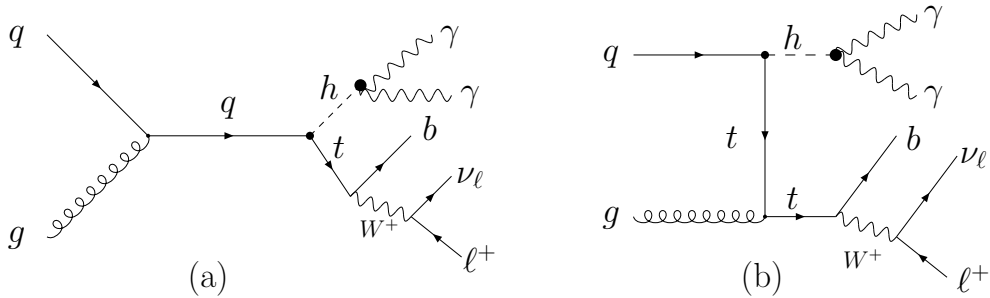}}
\vspace{-16cm}
\caption{The Feynman diagram for production of $th$ including the decay chain with leptonic top quark decay and
Higgs decay into diphoton.}
\label{qgaa}
\end{center}
\end{figure}

In this section, we perform the Monte Carlo simulation and explore the sensitivity of 14 TeV LHC to the top-Higgs FCNC couplings through the channel,
\begin{equation}\label{signal}
pp\to t(\to W^{+}b\to \ell^{+}\nu b)h(\to \gamma\gamma),
\end{equation}
where $\ell =e, \mu$. The Feynman diagram of
production and decay chain is presented in Fig.~\ref{qgaa}.

Obviously, the signal is taken as the single top plus a Higgs boson followed by the leptonic top quark decay and the Higgs boson decay into two photons, which is characterized by two photons appearing as a narrow resonance centered around the Higgs boson mass. The main SM backgrounds which yield the identical final states to the signal are $Whj$, $Wj\gamma\gamma$ and $tj\gamma\gamma$, where $j$ denotes non-bottom-quark jets. Besides, with fake photons due to misidentified jets or electrons, the reducible backgrounds such as $Wjj\gamma$, $tjj\gamma$, $tj\gamma$, $t\bar{t}\gamma$, $t\bar{t}\gamma\gamma$ can be important as well. On the other hand, the SM $thj$, $ZWh$, $W^{+}W^{-}h$ and $t\bar{t}h$ events can also be the sources of backgrounds for our signal. We have not included these backgrounds in the analysis due to very small cross sections. After including the
branching ratios and applying the cuts, a negligible number of events will survive.

All of these signal and backgrounds events are generated at LO using \texttt{MadGraph5-aMC$@$NLO} with the CTEQ6L PDF. \texttt{PYTHIA}~\cite{pythia} is utilized for parton shower and hadronization. \texttt{Delphes}~ \cite{delphes} is then employed to account for the detector simulations and \texttt{MadAnalysis5}~ \cite{ma5} for analysis, where the (mis-)tagging efficiencies and fake rates are assumed to be their default values, which is formulated as a function of the transverse momentum and
rapidity of the jets.  When generating the parton level events, we assume $\mu_R=\mu_F$ to be the default event-by-event value. The anti-$k_{t}$ algorithm~\cite{antikt} with the jet radius of 0.4 is used to reconstruct jets. The high order corrections for the dominant backgrounds are considered by including a $k$-factor, which is 1.12 for $Whj$~\cite{nlo-whj} and 1.3 for $Wj\gamma\gamma$~\cite{nlo-wjaa,14050301}, respectively. Here it should be mentioned that the $k$-factor for the LO cross section of $\sigma_{th}$ is chosen as 1.5 at the 14 TeV LHC~\cite{prd86-094014}. In order to avoid the double-counting issue of jets originated from matrix element calculation and the parton shower, we apply the MLM-matching implemented
in \texttt{MadGraph5-aMC$@$NLO}~\cite{MLM}. In practice, for the background events of $Whj$, we include both processes $pp \to Whj$ and $pp\to Whj+j$ to form an inclusive dataset,
and similarly for other backgrounds.

In our simulation, we generate $10^{6}$ events for the signals and backgrounds respectively. We first employ some basic cuts for the selection of events:
\beq
p_{T}^{j,b,\ell} &>& 25 \rm ~GeV, \quad |\eta_{j,b,\ell}|< 2.5, \nonumber \\
\slashed E_T^{miss} &>& 25 \rm ~GeV, \quad
\Delta R_{ij}>0.4 ~~(i,j = \ell, b, j, \gamma).
\eeq
where $p_{T}$ and $\eta$ are the transverse momentum and the pseudo-rapidity of jets and leptons while $\slashed E_T^{miss}$ is the missing transverse momentum.
  $\Delta R=\sqrt{(\Delta\phi)^{2}+(\Delta\eta)^{2}}$ is the particle separation with $\Delta\phi$ and $\Delta\eta$ being the separation in the azimuth angle and rapidity respectively.
For the signal, we require exactly one charged lepton, one $b$-jet, two photons and missing energy in the final state. To trigger the signal events, $N(\ell)=1$, $N(b)=1$ and $N(j)<2$
are applied, which can help to suppress the background events effectively, especially to the events with fake particles.

It should be mentioned that the $pp\to t\bar{t}\to thq$ process could also be considered as a source of top plus a Higgs boson if the light quark is missed by the detector.
It has been shown that this additional contribution is very significant for detecting the $\lambda_{tch}$ couplings due to the suppressed production cross section for the $cg\to th$ process~\cite{prd86-094014}. Therefore, we also consider this process when discussing the $tch$ couplings. In our calculation, the FCNC couplings are chosen to be
$\lambda_{tuh}=0.1$ and $\lambda_{tch}=0.1$, which are allowed by the low-energy experiments~\cite{prd81-077701,prd72-057504}.

\begin{figure}[htb]
\begin{center}
\vspace{-0.5cm}
\centerline{\epsfxsize=8cm\epsffile{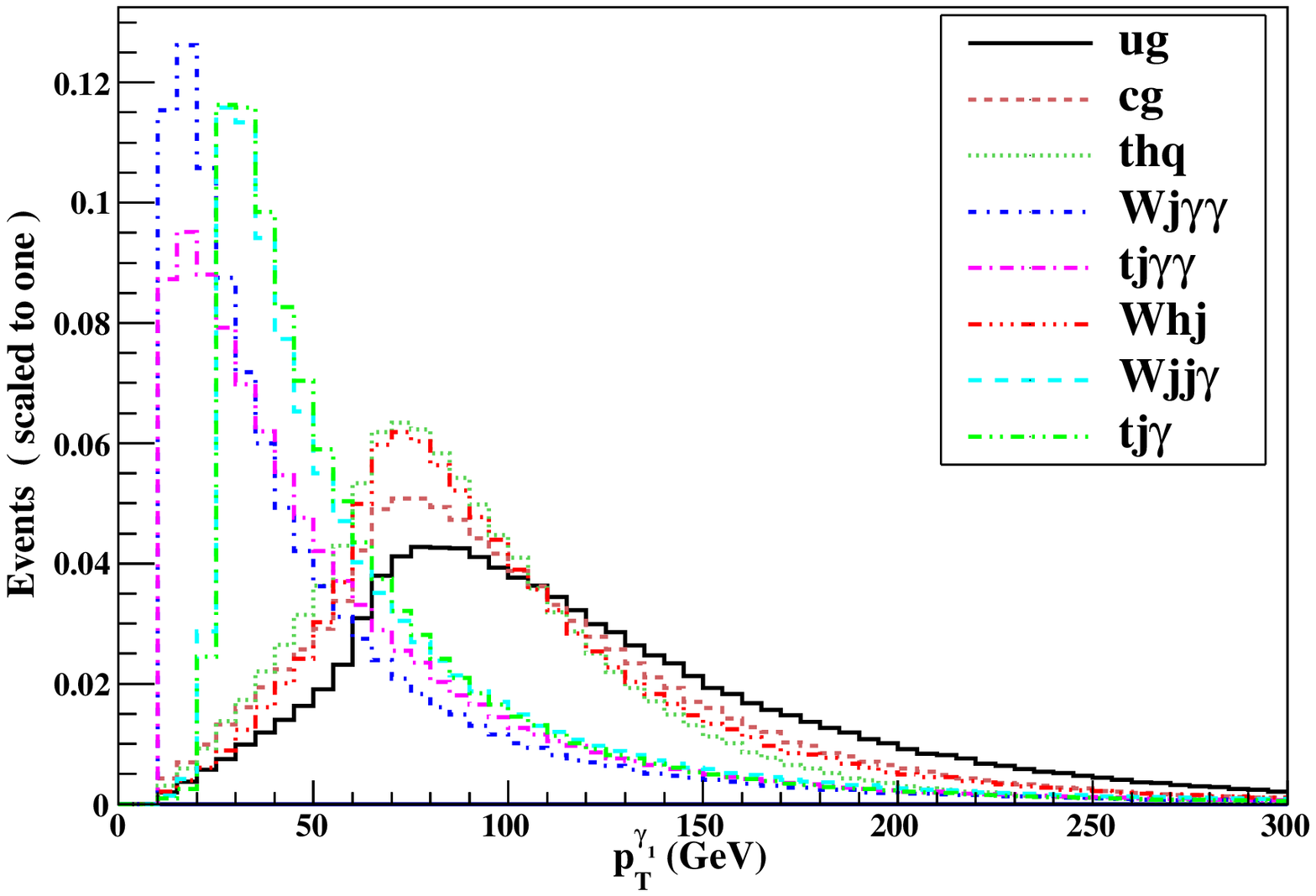}
\hspace{-0.5cm}\epsfxsize=8cm\epsffile{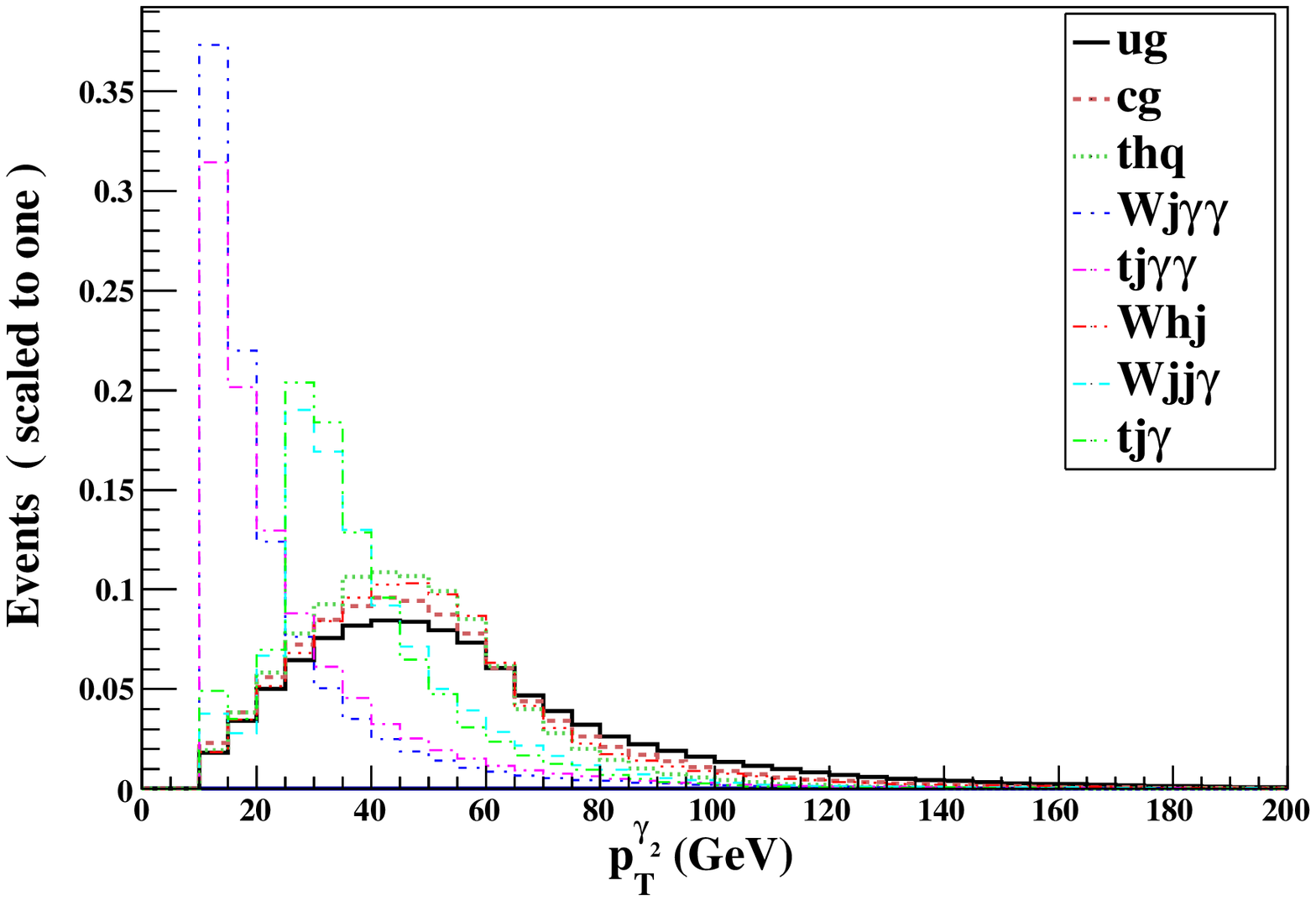}}
\caption{Normalized distributions of transverse momentum  $p_{T}^{\gamma_{1}}$ and $p_{T}^{\gamma_{2}}$ in the signals and backgrounds at 14 TeV LHC.}
\label{photon}
\end{center}
\end{figure}

In Fig.~\ref{photon}, we show the transverse momentum distributions of two photons in the signal and backgrounds at 14 TeV LHC. Since the two photons in the signal and the resonant backgrounds come from the Higgs boson, they have the harder $p_T$ spectrum than those in the non-resonant backgrounds.
Thus, we can apply the cuts $p_{T}^{\gamma_1}>55 \rm ~GeV$ and $p^{\gamma_2}_{T}>25 \rm ~GeV$ to suppress the non-resonant backgrounds.

\begin{figure}[htb]
\begin{center}
\vspace{-0.5cm}
\centerline{\epsfxsize=10cm \epsffile{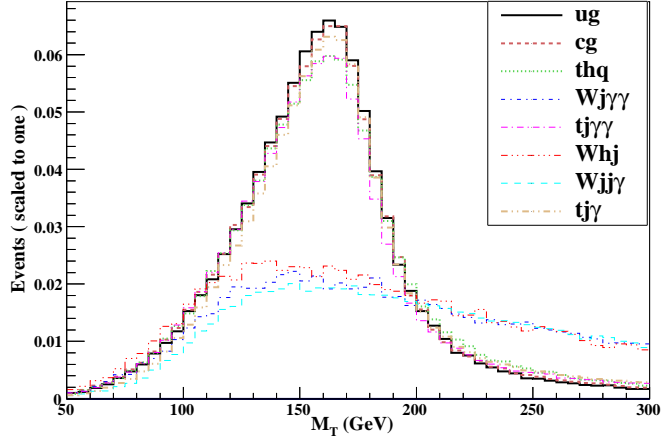}}
\caption{Normalized transverse mass distribution for the $b\ell \slashed E_T$  system at 14 TeV LHC.}
\label{mt}
\end{center}
\end{figure}

 Since there are only one $b$ jet and one lepton in the final states, it is easily to reconstruct the top quark transverse cluster mass, which
is defined as
\beq
M_T^{2}\equiv(\sqrt{(p_{\ell}+p_{b})^{2}+|\vec{p}_{T,\ell}+\vec{p}_{T,b}|^{2}}+|\vec{\slashed p}_T| )^{2}-|\vec{p}_{T,\ell}+\vec{p}_{T,b}+\vec{\slashed p}_T|^{2},
\eeq
where $\vec{p}_{T,\ell}$ and  $\vec{p}_{T,b}$ are the transverse momentums of the charged leptons and $b$-quark, respectively, and $\vec{\slashed p}_T$ is the missing transverse momentum determined by the negative sum of visible momenta in the transverse direction.
In Fig.~\ref{mt}, we show the transverse mass distribution for the $b\ell \slashed E_T$ system, which has been defined in the \texttt{MadAnalysis5}~\cite{epjc-74-3103}.
From this figure, we can see that the distributions of signal and backgrounds including top quark have peaks
around the top quark mass. Therefore, we choose the transverse mass $M_T$ cuts
\beq
120 \rm ~GeV< M_T < 190 \rm~GeV.
\eeq

\begin{figure}[htb]
\begin{center}
\vspace{-0.5cm}
\centerline{\epsfxsize=10cm \epsffile{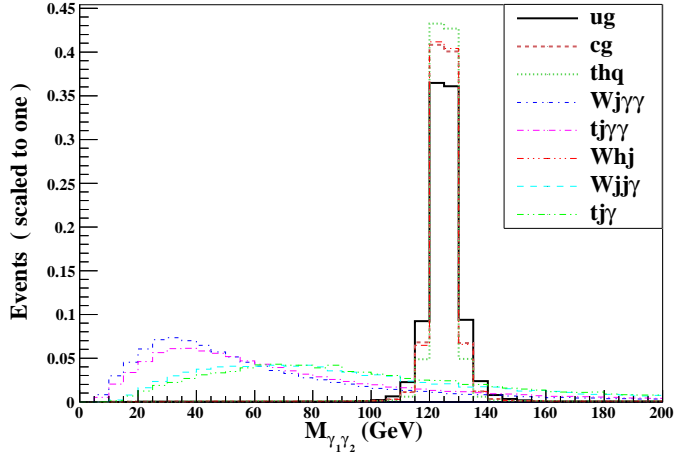}}
\caption{Normalized invariant mass distribution of two photons at 14 TeV LHC.}
\label{higgs}
\end{center}
\end{figure}
Next we consider to utilize the invariant mass distributions to further suppress the background. Fig.~\ref{higgs} illustrates the normalized invariant mass distribution $M_{\gamma\gamma}$ of the signal and backgrounds. We can see that the signals and the resonant backgrounds including the Higgs boson have peaks around 125 GeV. Thus we can further reduce the non-resonant backgrounds by the following cut:
\beq
120 \rm~GeV <M_{\gamma_{1}\gamma_{2}}<130 \rm ~GeV.
\eeq

For a short summary, we list all the cut-based selections here:
\begin{itemize}
  \item Basic cuts: $N(\ell) =1$, $N(b) =1$, $N(j)<2$, $p_T^{j,b,\ell}>25$ GeV, $|\eta_{j,b,\ell}|<2.5$, $\slashed E_T^{miss} > 25 \rm ~GeV$ and
      $\Delta R_{i,j}> 0.4 ~(i,j=j,b,\ell,\gamma)$;
 \item Cut 1 means the basic cuts plus $p_T^{\gamma_{1}}>55 \rm ~GeV$, $p_T^{\gamma_{2}}>25 \rm ~GeV$;
 \item Cut 2 means Cut 1 plus $120 \rm~GeV<M_T(\ell^{+} b \slashed E_T^{miss})< 190 \rm ~GeV$.

\item Cut 3 means Cut 2 plus requiring the invariant mass of the diphoton pair to be in the range $m_h\pm5$ GeV.
\end{itemize}

The cross sections of the signal and backgrounds after imposing
the cuts are summarized in Table~\ref{cutflow}. For the numbers of the cross sections
as listed in the Table~\ref{cutflow}, the FCNC couplings are chosen to be
$\lambda_{tuh}=0.1$ and $\lambda_{tch}=0.1$.
From Table~\ref{cutflow}, we can see that after all these cuts, all the backgrounds are suppressed efficiently and the total production cross section for the backgrounds is about $9.4\times 10^{-4}$ fb. The final production cross sections of $thj$, $t\bar{t}\gamma$ and $t\bar{t}\gamma\gamma$ are all at the level of $10^{-6}$ fb and thus they can be safely neglected. However, the cross section of the process $pp\to t\bar{t}\to thq$ is comparable to that of $cg\to th$.  As stated before, we should include these two processes when discussing the $tch$ couplings.
\begin{table}[htb]
\begin{center}
\caption{The cut flow of the cross sections (in $10^{-3}$ fb) for the signal and backgrounds at the 14 TeV LHC. The FCNC couplings are chosen to be
$\lambda_{tuh}=0.1$ and $\lambda_{tch}=0.1$. \label{cutflow}}
\vspace{0.2cm}
\begin{tabular}{|c|c|c|c|c|c|c|c|c|c|}
\hline
Cuts & $ug$ & $cg$ & $t\bar{t}\to thq$ & $Whj$ & $Wj\gamma\gamma$ & $tj\gamma\gamma$  & $Wjj\gamma$ & $tj\gamma$ & $tjj\gamma$  \\ \hline
Basic cuts & 69.2 & 10.1 & 9.6 & 0.57 & 270.5 & 3.6 & 2215 & 132 & 425 \\ \hline
Cut 1 & 56.5 & 7.45 & 7.5 & 0.41 & 35.5 & 5.3 & 16.8 & 5.54 & 2.72\\ \hline
Cut 2 &45.2&5.96&5.33&0.14&6.03&3.1&1.68&4.05&8.98\\ \hline
Cut 3& 33.9&4.95&4.53&0.12&0.54&0.015&0.09&0.13&0.05\\
\hline
\end{tabular} \end{center}\end{table}

In order to illustrate excluded detection potential regions of anomalous couplings to
reach a given statistical significance, we define the statistical significance (SS) as~\cite{ss}:
\beq
SS=\sqrt{2L[(S+B)\ln(1+\frac{S}{B})-S]},
\eeq
where $S$ and $B$ are the signal and background cross sections and $L$ is the integrated luminosity.
Here we define the discovery significance as $SS=5$ and exclusion limits as $SS=3$.
\begin{figure}[htb]
\begin{center}
\vspace{-0.5cm}
\centerline{\epsfxsize=8cm \epsffile{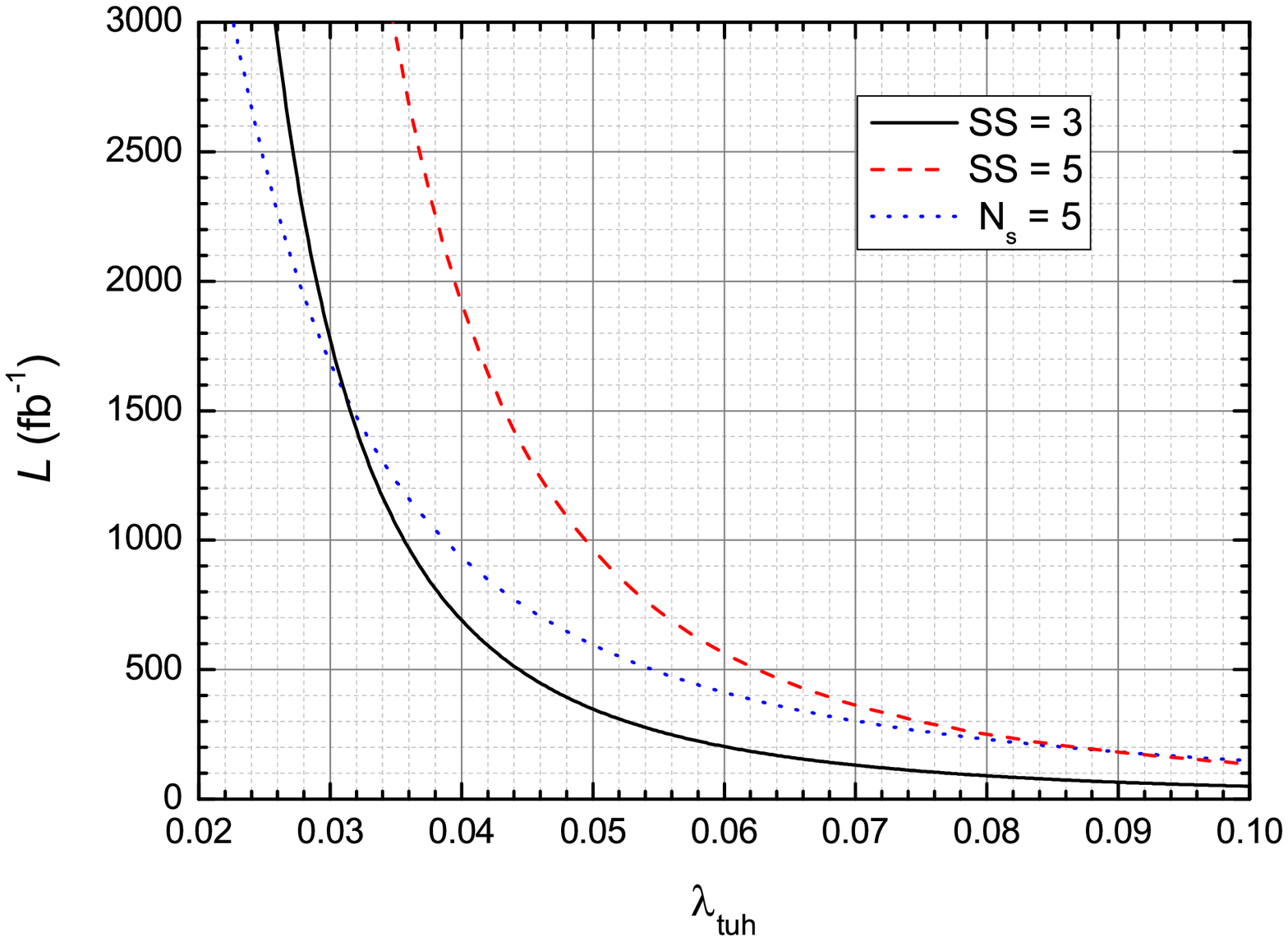}\epsfxsize=8cm \epsffile{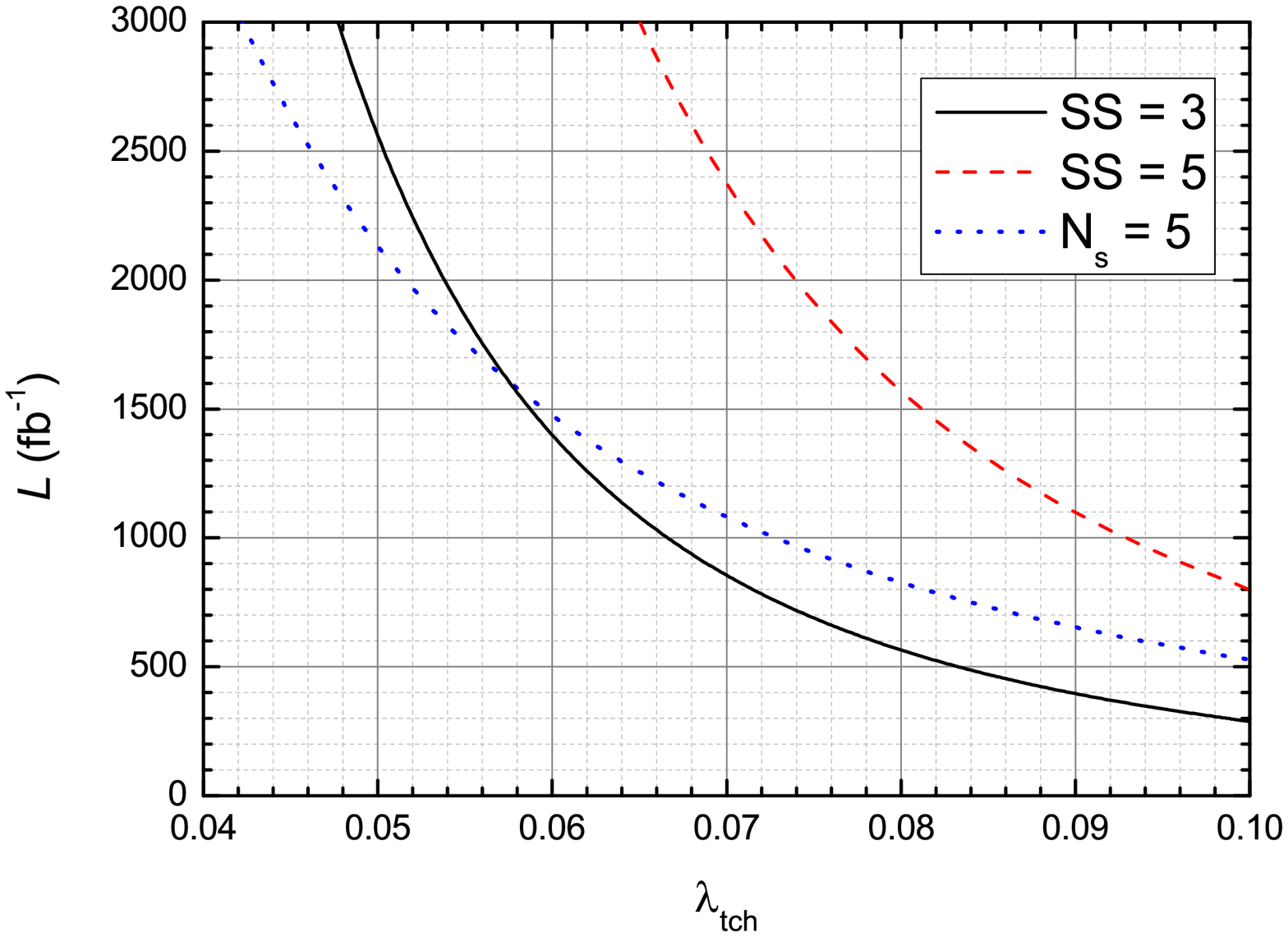}}
\caption{$3\sigma$ and $5\sigma$ contour plots for the signal in $L-\lambda_{tuh}$ (left) and $L-\lambda_{tch}$ (right) planes at 14 TeV LHC. The dotted line denotes the required minimum number of signal events $N_{s}=5$. }
\label{ss}
\end{center}
\end{figure}

In Fig.~\ref{ss}, we plot the excluded $3\sigma$ and $5\sigma$ discovery reaches in the plane of the integrated luminosity and the coupling parameter $\lambda_{tqh}$. To observe the signal at the LHC, we here also require the minimum 5 events for the signal. From Fig.~\ref{ss}, we can see that the $5\sigma$ C.L. discovery sensitivity of $\lambda_{tuh}$ is 0.062, 0.049 and 0.034 when the integrated luminosity is 500, 1000 and 3000 fb$^{-1}$, respectively. For the $thc$ couplings, the $5\sigma$ C.L. discovery sensitivity of $\lambda_{tch}$ is 0.065 when the integrated luminosity is 3000 fb$^{-1}$.

If no signal is observed, it means that the FCNC $tqh$ couplings can not be too large. The upper limits on the FCNC couplings $\lambda_{tuh}$ and $\lambda_{tch}$ can be respectively probed to 0.025 and 0.047 with $L =3000$ fb$^{-1}$.
These limits can be converted to the $3\sigma$ C.L. upper limits on the branching ratio $Br(t\to uh)=0.036\%$ and $Br(t\to ch)=0.13\%$, respectively. Compared with other phenomenological studies, we can see that our result is comparable with the sensitivity limits of LHC as $Br(t\to uh)<5\times 10^{-4}$ via multi-leptons channel with an integrated luminosity of
3000 $fb^{-1}$ at $\sqrt{s}=14$ TeV~\cite{13112028}.

\section{Charge ratio for signal and backgrounds}
As we discussed in Sec. II, the useful handle on tagging
$th$ production in searches with leptonic decays of top quark is the enhanced
abundance of positively charged leptons. Due to the difference between the
$u$-quark and $\bar{u}$-quark PDF, the cross sections of $th$ and $\bar{t}h$
are different for the processes $u(\bar{u})g\to t(\bar{t})h$ at the LHC.
Since the efficiencies of lepton selection and fake charged lepton
contamination are almost independent of charge, the top~(anti-top) asymmetry can be
directly translated in a corresponding lepton charge asymmetry.

\begin{table}[htb]
\begin{center}
\caption{The values of $R$ for the signal and backgrounds at the 14 TeV LHC. \label{ratio}}
\vspace{0.2cm}
\begin{tabular}{|c|c|c|c|c|c|c|c|c|c|c|}
\hline
\multirow{2}{*}{Process}& \multicolumn{3}{c|}{$ug(pp)\to th$}&\multirow{2}{*}{$cg$} & \multirow{2}{*}{$Whj$} & \multirow{2}{*}{$Wj\gamma\gamma$} & \multirow{2}{*}{$tj\gamma\gamma$} & \multirow{2}{*}{$Wjj\gamma$} & \multirow{2}{*}{$tj\gamma$} \\ \cline{2-4}
 &{$\mu_0/2$}&{$\mu_0$}&{$2\mu_0$} &&&&&& \\\cline{1-10}
\hline
$R$ &6.76(4.74)&6.75(4.67)&6.75(4.64)&1.0&1.57&1.17&1.59&1.17&1.45\\
\hline
\end{tabular} \end{center}\end{table}

\begin{figure}[htb]
\begin{center}
\vspace{-0.5cm}
\centerline{\epsfxsize=9cm\epsffile{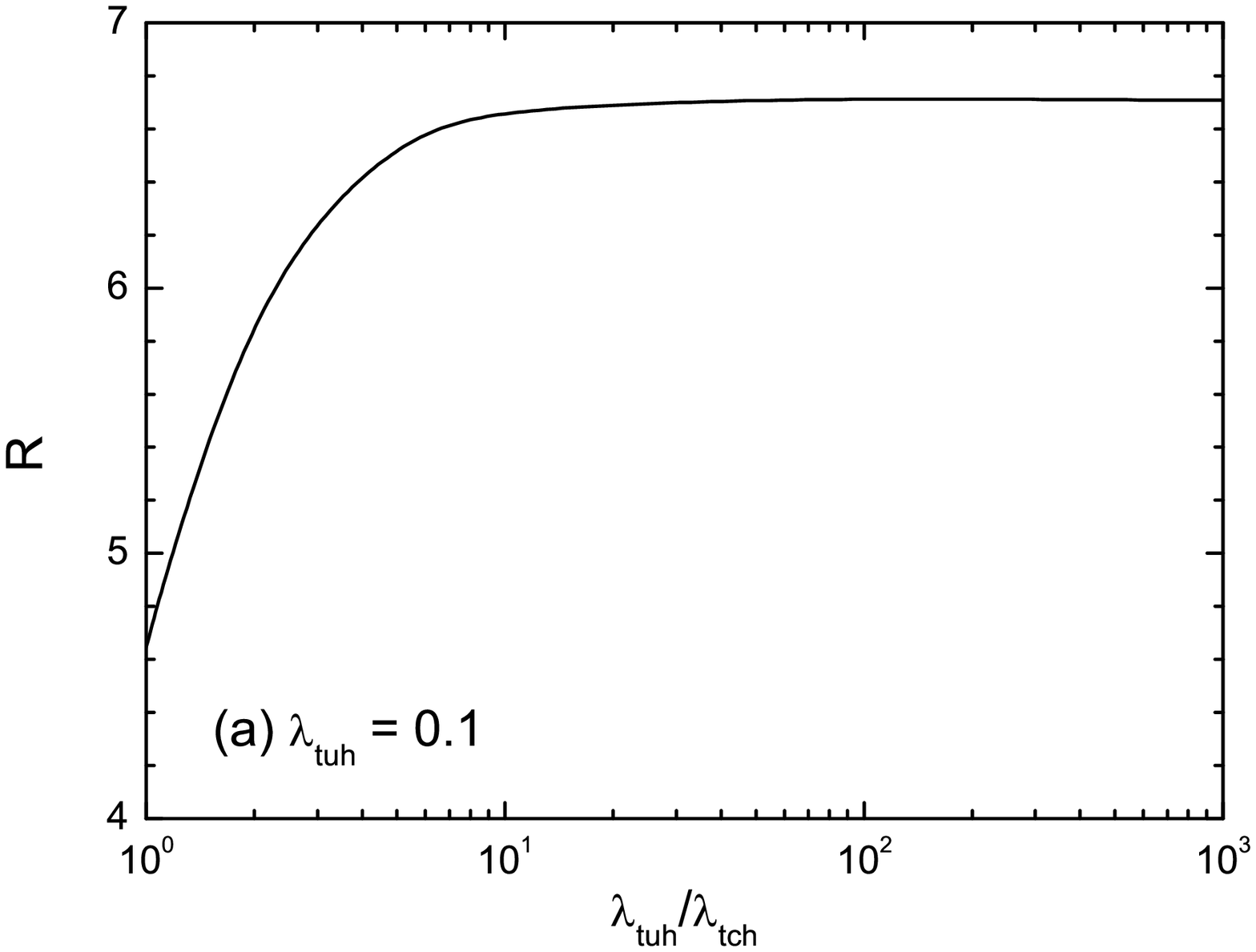}
\hspace{-0.5cm}\epsfxsize=9cm\epsffile{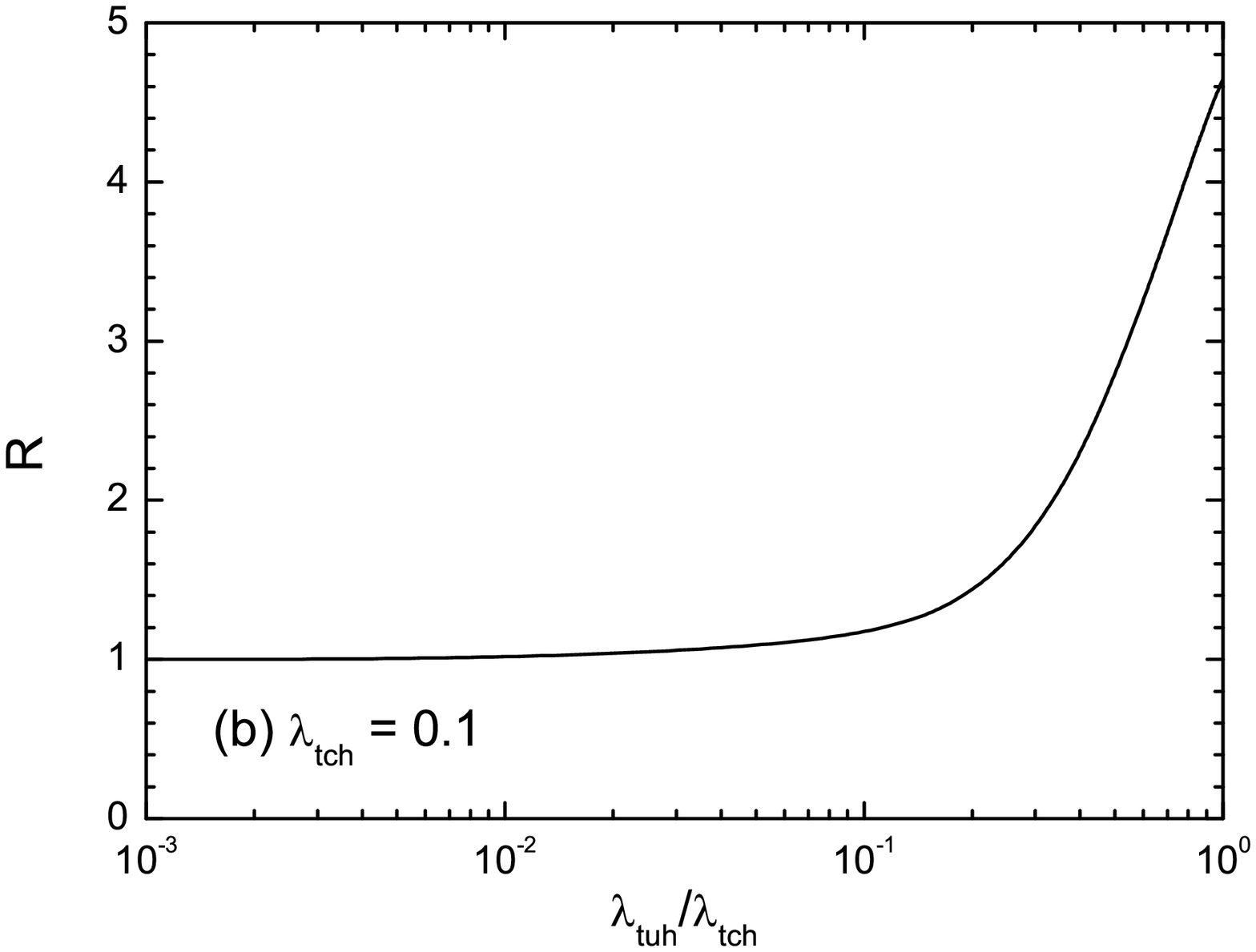}}
\caption{The dependence of the charge ratio $R$ for the process $pp\to th$ on the various of ratio values $\lambda_{tuh}/\lambda_{tch}$ at the 14 TeV LHC with (a) $\lambda_{tuh}=0.1$ and (b) $\lambda_{tch}=0.1$.}
\label{r}
\end{center}
\end{figure}

 In this analysis, we define a ratio $R=N^{+}/N^{-}$ as
the number of events with positive charged lepton to the
number of events with negative charge. For the signal and relevant backgrounds, the values of $R$ are listed in Table~\ref{ratio} at 14 TeV LHC with the same basic cuts.
Here we present three typical renormalization scale and factorization scale as $\mu_R=\mu_F=\mu_0/2$, $\mu_0$ and $2\mu_0$, respectively. One can see that, for $\lambda_{tch}=\lambda_{tuh}=0.1$, the value of the charge ratio for the signal is insensitive to the renormalization and factorization scales, and it is significantly larger than that for the SM backgrounds. In Fig.~\ref{r} we show the dependence of the charge ratio $R$ for the process $pp\to th$ on the various of ratio values $\lambda_{tuh}/\lambda_{tch}$ at the 14 TeV LHC. One can see that for the case of $\lambda_{tuh}\gg \lambda_{tch}$, the contribution from the $tuh$ vertex can enhance the charge ratio, while for the case of  $\lambda_{tuh}\ll \lambda_{tch}$, the contribution from the $tch$ vertex can significantly dilute the ratio.
Therefore, studying the charged ratio for lepton can be not only used to
discriminate between signal and backgrounds, but also used to determine that the signal comes from the up initiated production channel and charm initiated production channel.

\section{CONCLUSION}
In the work, we investigated the process $pp \to th$ induced by the top-Higgs FCNC couplings at the LHC. We also studied the observability of top-Higgs FCNC couplings through the process $pp\to t(\to W^{+}b\to \ell^{+}\nu b)h(\to \gamma\gamma)$ and proposed the charge ratio of signal and backgrounds of the charge lepton. From our numerical calculations and the phenomenological analysis we found the following points:
\begin{enumerate}
\item
The cross section of $pp\to ug \to th$ is larger than that for other process due to the larger PDF of the $u$-quark, which means that
the sensitivity to the FCNC coupling $\lambda_{tuh}$ is better than $\lambda_{tch}$.

\item
We further studied the observability of top-Higgs FCNC couplings through the process $pp\to t(\to W^{+}b\to \ell^{+}\nu b)h(\to \gamma\gamma)$ and found that in some parameter regions, the LHC may observe the above signals at the $5\sigma$ confidence level. Otherwise, the branching ratios $Br(t\to uh)$ and $Br(t\to ch)$ can be respectively probed to $0.036\%$ and $0.13\%$ at $3\sigma$ level at 14 TeV LHC with the integrated luminosity of 3000 fb$^{-1}$.

\item
The charge ratio for the lepton in the signal is significantly larger than that for the SM backgrounds, which can be not only used to
discriminate between signal and backgrounds, but also used to determine that the signal comes from the up initiated production channel and charm initiated production channel.
\end{enumerate}

\begin{acknowledgments}
This work is supported by the National Natural Science
Foundation of China under the Grant No.11235005, the Joint Funds
of the National Natural Science Foundation of China (U1304112) and the Foundation of He'nan Educational Committee (2015GGJS-059).
\end{acknowledgments}

\end{document}